# Distributed Double Auctions for Large-Scale Device-to-Device Resource Trading

Shuqin Gao
Singapore University of Technology and Design
shuqin_gao@mymail.sutd.edu.sg

Costas Courcoubetis
Singapore University of Technology and Design
costas@sutd.edu.sg

Lingjie Duan
Singapore University of Technology and Design
lingjie_duan@sutd.edu.sg

## ABSTRACT

Mobile users in future wireless networks face limited wireless resources such as data plan, computation capacity and energy storage. Given that some of these users may not be utilizing fully their wireless resources, device-to-device (D2D) resource sharing is a promising approach to exploit users' diversity in resource use and for pooling their resources locally. In this paper, we propose a novel two-sided D2D trading market model that enables a large number of locally connected users to trade resources. Traditional resource allocation solutions are mostly centralized without considering users' local D2D connectivity constraints, becoming unscalable for large-scale trading. In addition, there may be market failure since selfish users will not truthfully report their actual valuations and quantities for buying or selling resources. To address these two key challenges, we first investigate the distributed resource allocation problem with D2D assignment constraints. Based on the greedy idea of maximum weighted matching, we propose a fast algorithm to achieve near-optimal average allocative efficiency. Then, we combine it with a new pricing mechanism that adjusts the final trading prices for buying and selling resources in a way that buyers and sellers are incentivized to truthfully report their valuations and available resource quantities. Unlike traditional double auctions with a central controller, this pricing mechanism is fully distributed in the sense that the final trading prices between each matched pair of users only depend on their own declarations and hence can be calculated locally. Finally, we analyze the repeated execution of the proposed D2D trading mechanism in multiple rounds and determine the best trading frequency.

## 1 INTRODUCTION

### 1.1 Background and Motivation

The recent growth of mobile devices and applications has been unprecedented as smartphones become extremely popular in our daily life. Future wireless networks are challenged by the different quality of service requirements from a wide variety of new emerging mobile applications, such as video streaming, face recognition, natural language processing, etc. These applications demand not only ubiquitous high-speed wireless access and fast response time, but also intensive computation and energy consumption. However, mobile users only have limited wireless resources such as date plans, energy storage and computation capacity to serve these applications. Given that a number of these users may not be utilizing fully their wireless resource, device-to-device (D2D) resource trading[1] among neighboring mobile users is envisioned as a promising approach that improves network utilization and reduces congestion and latency by exploiting the diversity over time in resource use of the mobile users. Moreover, with the technological advancements in smartphones, today D2D resource trading can be easily implemented. For examples, an iPhone or Android phone can easily open up personal hotspot and share data connects to another device in the vicinity [17]; and many Huawei and Samsung phones can wireless charge others. Other than sharing data plan, power and edge computing capacity in wireless networks, a user can also cache popular files and transfer to her neighbors via local links after installing some customized apps in smartphone [5].

Some recent works have been carried out on D2D resource sharing aiming at improving efficiency in using scarce resources such as data plan, power, computation capacity and cache memory in wireless networks. In [17, 19], the secondary markets for demand-heterogeneous users to trade their monthly data plans via personal hotspots (PHs) or other centralized platforms are proposed and analyzed. [4] allows users to perform D2D power cooperation by transmitting their spare power to other nearby users. A novel D2D mobile task offloading framework is developed in [15] to enable users to share the computation resources among each other. [5] investigates the cooperative local caching under heterogeneous file preferences, and [16] studies the incentives for nearby mobile users to cooperatively downloading video segments. In [10, 11], the sharing platforms are designed to collect the social information from users when traveling and benefit all users that participate.

However, most of the existing research on D2D resource sharing focuses on centralized resource allocation and does not address scenarios involving a large number of mobile users. Communication and computation overheads make it difficult to determine resource allocations and participant compensations on a global scale. A challenge remains on how to design scalable and efficient allocation algorithms based on users' local information only. Furthermore, users being selfish and non-cooperative in nature, are unwilling to truthfully report their private information to the allocation algorithm. This challenges us even further to design the appropriate incentive structure that complements the distributed allocation algorithms and makes users interact truthfully with the system [1].

This work was supported by the Ministry of Education, Singapore, under its Academic Research Fund Tier 2 Grant (Project No. MOE2016-T2-1-173).

[1]In our case D2D refers to interactions between wireless devices that are connected in a one-hop way via a local wireless link. We use the term 'trading' instead of just 'sharing' to emphasize the monetary incentives underlying such resource sharing.

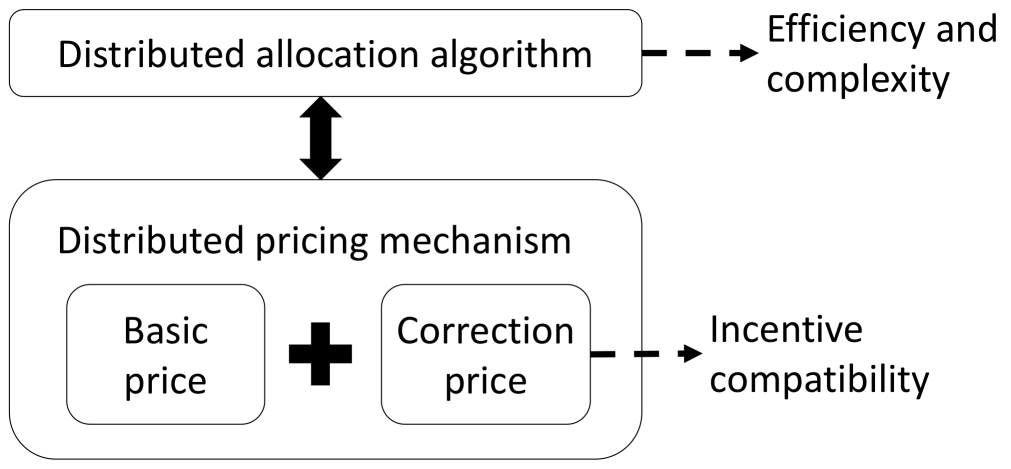

**Figure 1: Our double auction mechanism. It includes a distributed allocation algorithm that balances efficiency with complexity and a distributed pricing mechanism for ensuring incentive compatibility and participation.**

## 1.2 Our Solution and Contributions

In this work, we design a novel two-sided market model for large-scale D2D trading, where mobile users trade resources with each other in proximity via local wireless links (e.g., short-range communications). We study a challenging scenario that during each trading round, there exist a large number of users who want to trade with each other as consumers or suppliers of resources. Their geographic locations determine who to trade with whom potentially locally and the D2D assignment constraints. To achieve efficient resource trading, we propose a double auction mechanism to optimally allocate resources using private information from both buyers and sellers (i.e., the unit value and demanded quantity as a buyer, or the unit cost and supplied quantity as a seller). Double auction is widely used for two-sided market model with multiple buyers and multiple sellers [12, 18]. Though some recent works study how to implement double auction in a decentralized system (e.g., [7, 8]), they still need a central controller to gather all users' information and are not suitable for a large-scale D2D trading scenario. Differently, our proposed distributed double auction mechanism nicely has all the following properties: (i) fully distributed operation without a central controller, (ii) high efficiency and low complexity to solve the constrained resource allocation problem, and (iii) incentive compatibility to elicit truthful private information from users. We believe that our solution is useful enough to make it practically interesting in a variety of contexts. The mechanism consists of a distributed allocation algorithm and a distributed pricing mechanism as shown in Fig. 1.

Firstly, we investigate the resource allocation problem between buyers and sellers with practical D2D assignment constraints. Most existing methods to optimally solve this problem require a central controller to gather all participants' information and perform the computation centrally. The only known distributed algorithm is in [3]. But to find the optimum it requires a prohibitively high average computational complexity when the network is large. To reduce this complexity, we propose a 2-approximation distributed allocation algorithm based on the greedy idea of maximum weighted matching in [6, 14]. The proposed algorithm has only linear complexity and is further asynchronous for the purpose of easy implementation. Numerical results show that the proposed algorithm has near-optimal average performance at a significantly lesser running time as compared to the optimal algorithm mentioned above, as tested in our large-scale network simulations.

Secondly, we design a novel distributed pricing mechanism to elicit truthful private information from users. There are some known mechanisms that can be used in double auction design and are incentive compatible, such as the Vickrey-Clarke-Groves (VCG) and Arrow-d'Aspremont-Gerard-Varet (AGV). In VCG, reporting the true information is a dominant strategy [13], but it requires a *central controller* to decide the price for each transaction and has high computational complexity. AGV [2], besides requiring centralized computation like VCG, it also requires that once users have accepted to join the trading platform, they cannot strategically choose which rounds to participate in the trade based on their private information. This assumption of participating commitment is hard to meet in practice, making AGV not a viable alternative.

Our novel trading price design uses two components: a basic price component and a correction price component (see Fig. 1). Using only the first component (which corresponds to the mid value between the buyer's reported value and the seller's reported cost), all users have positive gain by participating, the system is budget balanced, but it is not incentive compatible. The second component acts as a correction to the first and induces incentive compatibility. By using this pricing mechanism, a positive utility is still guaranteed in each round for any user no matter if she participates as a buyer or a seller. Thus, our mechanism is also individually rational. Furthermore, to recover the correction price paid to all users and keep the budget balanced, the mechanism requires the platform to charge users a subscription fee (say per month). Our main contributions are:

- We propose a 2-approximation distributed allocation algorithm for the resource allocation problem with unique D2D assignment constraints. This algorithm has near-optimal average performance at a significantly lesser running time as compared to the optimal benchmark.
- We complement the allocation algorithm with a novel pricing mechanism. The resulting double auction mechanism can be nicely implemented in *fully distributed environments* and is *incentive compatible, individually rational and becomes ex-ante budget balanced* by charging users a subscription fee.
- We further extend the proposed mechanism to multiple rounds and study the best trading frequency over time by balancing trading opportunity and waiting cost of users. We also show that the proposed allocation algorithm has another advantage. When many of the users remain active during multiple rounds, it keeps existing trading pairs unchanged during each round, while the optimal allocation algorithm creates different trading pairs with high switching cost.

The paper is organized as follows. In Section 2 we present our model of the market place and the basic structure of the proposed trading mechanism. In Section 3 we propose a new distributed algorithm for matching supply and demand that is fast and efficient. In Sections 4 and 5 we propose the pricing algorithm with the price correction that makes it incentive compatible. Finally, in Section 6 we analyze the performance of repeated trading and end with some conclusions.

## 2 SYSTEM MODEL

### 2.1 Problem Description

We propose a D2D trading market model for a large number of potential users to trade resources with each other in proximity via local wireless links (e.g., short-range communications). In this market trade takes place continuously in repeated trading rounds and is supervised by the 'system platform' to which users must enroll in order to obtain the right to trade. The challenge we address in this research is how to implement the trading functionality of this platform using distributed operation.

At the beginning of a trading round, an individual user chooses to participate as a buyer if she is willing to pay for using more resources for utility or as a seller if she has excess resources to share for profit. Thus, the participating users of each round belong to two groups: the buyer group $\mathcal{M} = \{1, 2, \cdots, M\}$ and the seller group $\mathcal{N} = \{1, 2, \cdots, N\}$. According to the users' current locations in this round, each buyer $i \in \mathcal{M}$ is only within the coverage of a subset of nearby sellers, denoted as $\mathcal{S}(i)$, and each seller $j \in \mathcal{N}$ is able to serve a subset of buyers within her coverage area, denoted as $\mathcal{B}(j) = \{i \in \mathcal{M} : j \in \mathcal{S}(i)\}$. Note that the D2D trading market model (including groups $\mathcal{M}$ and $\mathcal{N}$) changes over time according to users' random arrival, movement and departure which will be detailed later in Section 6. There we will apply our distributed double auction mechanism for each trading round, and show the parameter design (e.g., trading frequency) for such dynamic trading over time.

In any given round, we view the market model as an instance of a random bipartite matching graph $G = (M, N, \{\mathcal{S}(i)\}, \{\mathcal{B}(j)\})$ where buyers and sellers are nodes, the wireless links are edges, and $G$ have a certain distribution. A simple interpretation of the model is that a typical user, when participating, corresponds to a random node in $G$. She acts at different times as a buyer or a seller and her connectivity with the rest of the nodes is chosen at random according to the distribution of $G$. Before she starts to trade, she already knows all her neighbors via the local communication. A small-scale illustrative instance of the D2D trading market is shown spatially on the ground in Fig. 2, which can be translated to the bipartite matching graph between $M = 6$ buyer nodes and $N = 4$ seller nodes.

Many wireless resources (e.g., data plan, power, computation capacity and cache memory) are divisible and one can aggregate resources from different sources to use at the same time (e.g., [17], [16]). In our model, the traded resources are assumed to be measured in certain commonly agreed units by the market participants. For a market of data plan allowance, a unit might correspond to a megabyte of data, or, for a market of cache memory, a unit might correspond to a standard size file. Moreover, we consider that the demand of a single buyer can be served by aggregating resources from multiple sellers. Hence, we generally allow each buyer $i$ to *buy resources from multiple sellers* to meet her demand $\alpha_i$ and each seller $j$ also to *serve more than one buyer at one time* to sell her supply $\beta_j$. We also consider that buyers and sellers have linear value and cost functions[2], respectively, for the amount of resources they trade. The value of obtaining a unit resource is denoted by $v_i$ for buyer $i$

[2]If nonlinear, we can still apply linear approximation to obtain linear terms for users' values and costs.

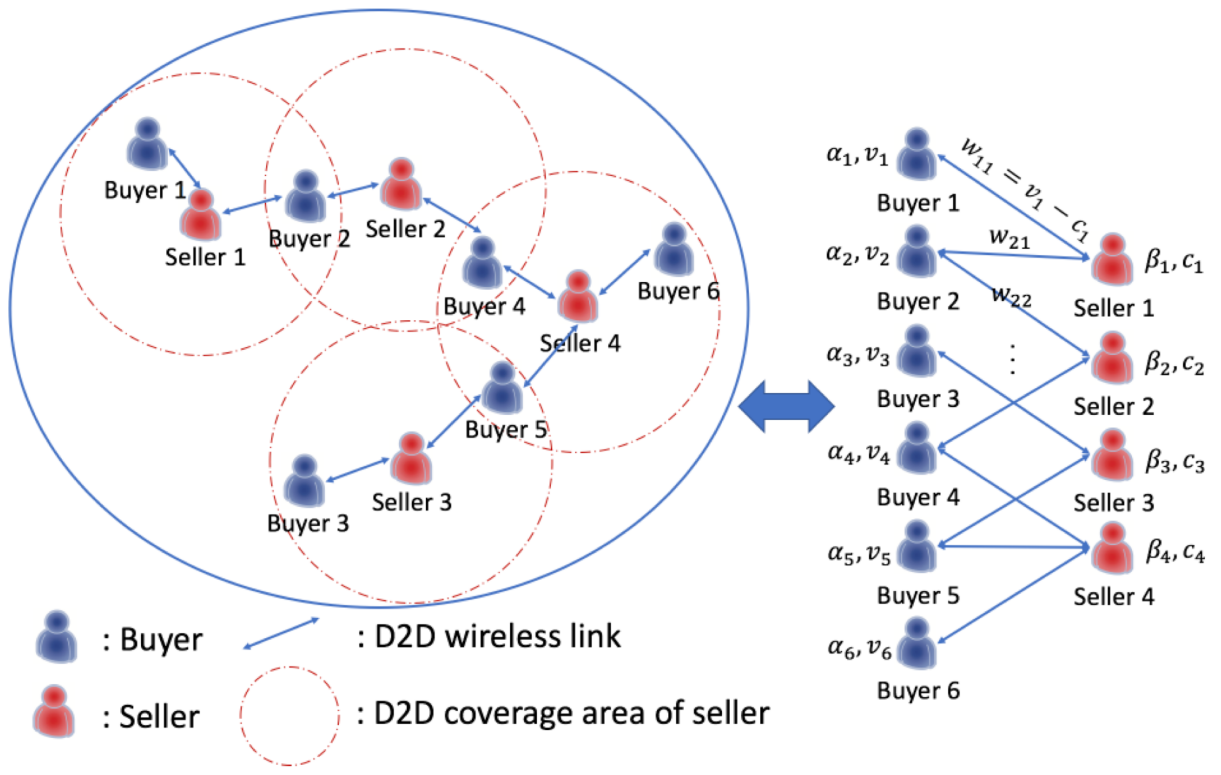

**Figure 2: A illustrative instance of the D2D trading market with $M = 6$ buyers and $N = 4$ sellers is captured spatially on the left-hand-side subfigure. Then we can abstract their local connectivity as the bipartite matching graph on the right-hand-side subfigure.**

and the cost incurred by offering a unit resource is denoted by $c_j$ for seller $j$.

In each trading round, the aim of the system platform is to operate efficiently and maximize the difference between the total value created to buyers by the allocated resources and the total cost caused by depriving these resources from the users that sell them, known as the social welfare. Specifically, the social welfare maximization problem in our market can be formulated as a network flow problem from the bipartite matching graph as follows:

$$\mathcal{P}_1: \quad \max \sum_{i \in \mathcal{M}} \sum_{j \in \mathcal{S}(i)} w_{ij} f_{ij}, \tag{1a}$$

$$\text{s.t.} \sum_{j \in \mathcal{S}(i)} f_{ij} \le \alpha_i, \quad \forall i \in \mathcal{M}, \tag{1b}$$

$$\sum_{i \in \mathcal{B}(j)} f_{ij} \le \beta_j, \quad \forall j \in \mathcal{N}, \tag{1c}$$

$$f_{ij} \in \{0, 1, 2, \cdots\}, \forall i \in \mathcal{M}, j \in \mathcal{S}(i), \tag{1d}$$

where $w_{ij} = v_i - c_j$ in (1a) and Fig. 2 denotes the net benefit of a unit resource allocation between buyer $i$ and seller $j$, and $f_{ij}$ is the optimization variable denoting the amount of divisible resource units allocated to pair $(i, j)$. The total amount of resources allocated to buyer $i$ is constrained by $\alpha_i$ according to the demand constraint in (1b), and the constraint in (1c) tells that seller $j$ cannot give out more than she owns. Constraint in (1d) ensures that each buyer $i$ can only be assigned to seller $j$ in set $\mathcal{S}(i)$, which we refer to as a D2D assignment constraint.

### 2.2 Distributed Double Auction Mechanism

First, note that, in our decentralized D2D trading market with many buyers and sellers, problem $\mathcal{P}_1$ cannot be solved centrally due to the high communication and computation overhead. Second, users are selfish and unwilling to truthfully report private information, which introduces an information challenge for solving problem $\mathcal{P}_1$ requiring full knowledge of demand $\alpha_i$ and value $v_i$ (supply $\beta_j$ and cost $c_j$) from each buyer $i$ (seller $j$). However, these parameters are known to individual users at the beginning of the trading round and are private information unknown to others. If these users misreport this information, even if problem $\mathcal{P}_1$ is solved optimally, the value of the actual social welfare achieved might be far from the true optimum resulting to a market failure.

To handle these two challenges, we need to develop a distributed truthful double auction mechanism to determine how to match buyers and sellers and how to allocate and pay/charge resources depending on the users' declarations. Let $(\hat{\alpha}_i, \hat{v}_i)$ and $(\hat{\beta}_j, \hat{c}_j)$ denote the declarations of buyer $i$ and seller $j$, respectively, regarding their private information. There are two key components in our mechanism design: the *allocation algorithm* to determine the resource allocation $f_{ij}$ by solving problem $\mathcal{P}_1$ and the *pricing mechanism* to determine the unit buying price $p_{ij}^B$ paid by buyer $i$ and the unit selling price $p_{ij}^S$ paid to seller $j$ for each matched pair $(i, j)$ with allocation $f_{ij} > 0$. In each round, the double auction mechanism runs by the following procedure:

- At the beginning, each buyer $i$ submits her declaration $(\hat{\alpha}_i, \hat{v}_i)$ to all the neighboring sellers belonging to $\mathcal{S}(i)$ and each seller $j$ also submits her declaration $(\hat{\beta}_j, \hat{c}_j)$ to all the neighboring buyers belonging to $\mathcal{B}(j)$.
- We run a distributed allocation algorithm to match buyers and sellers and for each matched pair $(i, j)$, determine the corresponding allocation $f_{ij}$ depending on local[3] declarations.
- For each matched pair $(i, j)$, we run a distributed pricing mechanism to determine the final buying price $p_{ij}^B$ and the final selling price $p_{ij}^S$ that are only based on the declarations of buyer $i$ and seller $j$.

The rest of the paper deals with developing the above mechanisms. The desired properties of the pricing mechanism are:

- (E1) Incentive compatibility. The users are induced to truthfully report their private information (i.e., $\hat{\alpha}_i = \alpha_i, \hat{v}_i = v_i, \hat{\beta}_j = \beta_j, \hat{c}_j = c_j, \forall i, j$).
- (E2) Individually rationality. A user should not obtain negative utility from participating in each round of the auction.
- (E3) Ex-ante budget balance. In the long run, the total amount of money collect from users should be no less than the amount paid to users. This ensures viability of the system platform in the long run.
- (E4) Long-term participation. A user that pays the subscription fee to the platform still expects a non-negative average total profit.

## 3 DISTRIBUTED ALLOCATION ALGORITHM

In this section, we propose and analyze a distributed allocation algorithm to solve problem $\mathcal{P}_1$ under complete information. We will deal with incentive compatibility to ensure complete information in the next sections.

There have been distributed greedy algorithm proposals (e.g., [6]) to solve the single-unit version of problem $\mathcal{P}_1$, which is known as the maximum weighted matching problem. In this problem, each buyer (seller) demands (supplies) a unit resource. Simple extension to multiple units by providing multiple copies of buyers and sellers fails because of greatly increasing the dimensionality of the problem (the network size increasing significantly from $M + N$ to $\sum_{i \in \mathcal{M}} \alpha_i + \sum_{j \in \mathcal{N}} \beta_j$) and also because they introduce competition between copies of the same buyer (seller). We address these issues by proposing Algorithm 1 in the following.

**Algorithm 1:** Multi-unit weighted matching for solving problem $\mathcal{P}_1$

**Initialization:** $f_{ij} = 0, \forall i \in \mathcal{M}, j \in \mathcal{S}(i)$.

In each iteration, repeat the following two phases:

**Requesting phase:**

For each buyer $i \in \mathcal{M}$ with unsatisfied demand $\alpha_i > 0$:

- Find $\alpha_i$ neighboring sellers with the largest weights from the neighbor set $\mathcal{S}(i)$ and sort them in a non-increasing weight order, denoted as $\{j_1, j_2, \cdots, j_{\min\{\alpha_i, |\mathcal{S}(i)|\}}\}$ with $w_{ij_1} \geq w_{ij_2} \geq \cdots \geq w_{ij_{\min\{\alpha_i, |\mathcal{S}(i)|\}}}$.
- From $k = 1$ to $k = \min\{\alpha_i, |\mathcal{S}(i)|\}$, compute the allocation requested one by one as follows:

$$f_{ij_k}^B = \begin{cases} \min\{\alpha_i, \beta_{j_k}\}, & \text{if } k = 1, \\ \min\{\alpha_i - \sum_{t=1}^{k-1} \beta_{j_t}, \beta_{j_k}\}, & \text{if } k \geq 2. \end{cases} \quad (2)$$

  By doing so, buyer $i$ allocates her demand $\alpha_i$ greedily to her neighboring sellers according to the non-increasing weight order until her demand is fully met.
- Send a request for $f_{ij_k}^B$ units of resources to each seller $j_k$ if $f_{ij_k}^B > 0$.

For each seller $j \in \mathcal{N}$ with leftover supply $\beta_j > 0$:

- Seller $j$ applies the 'symmetric' procedure presented above to compute $f_{ij}^S$ for all her neighboring buyers $i \in \mathcal{B}(j)$ sorted similarly.

**Assignment Phase:**

For each pair $(i, j)$ requested by both buyer $i$ and seller $j$, i.e., $f_{ij}^B > 0$ and $f_{ij}^S > 0$:

- Update allocation $f_{ij} = f_{ij} + \min\{f_{ij}^B, f_{ij}^S\}$, demand $\alpha_i = \alpha_i - \min\{f_{ij}^B, f_{ij}^S\}$ and supply $\beta_j = \beta_j - \min\{f_{ij}^B, f_{ij}^S\}$.
- If unsatisfied demand $\alpha_i = 0$, remove buyer $i$ from the neighbor sets of all her neighboring sellers by updating $\mathcal{B}(j') = \mathcal{B}(j') \setminus \{i\}, \forall j' \in \mathcal{S}(i)$.
- If leftover supply $\beta_j = 0$, remove seller $j$ from the neighbor sets of all her neighboring buyers by updating $\mathcal{S}(i') = \mathcal{S}(i') \setminus \{j\}, \forall i' \in \mathcal{B}(j)$.

Algorithm 1 is clearly *distributed.* Each user repeats the steps of the algorithm based on local information and will stop once her demand is met (or supply is sold out) or sees no available neighbor. She does not need to wait for the 'global' termination of the distributed algorithm, i.e., all users reaching a termination condition. Global termination is ensured within $\min\{\sum_{i \in \mathcal{M}} \alpha_i, \sum_{j \in \mathcal{N}} \beta_j\}$ iterations since at least one unit resource is added to the existing allocation in each iteration. Note that once a unit resource is added to the existing allocation, it cannot be removed later. Moreover, in each iteration, an unsatisfied buyer $i$ (or a seller $j$ with leftover supply) needs to update her allocation requests only when the allocation corresponding to her neighbors in $\mathcal{S}(i)$ (or $\mathcal{B}(j)$) changes in last iteration. The allocation corresponding to buyer $i$ (or seller $j$) can change at most demand $\alpha_i$ (or supply $\beta_j$) times.

[3] We use the term 'local' to refer to information or actions involving neighboring nodes.

Therefore, this algorithm runs in linear $O(\sum_{i\in\mathcal{M}}|\mathcal{S}(i)|)$ time. Note that $\sum_{i\in\mathcal{M}}|\mathcal{S}(i)| = \sum_{j\in\mathcal{N}}|\mathcal{B}(j)|$ and is actually the total number of edges in the bipartite matching graph. The next proposition summarizes the properties of our Algorithm 1. We will complement the proposed allocation algorithm with a novel pricing mechanism later in Section 5.1 and such monotonicity properties are needed for the incentive compatibility proofs there.

Proposition 1. *Algorithm 1 achieves an approximation ratio of $\frac{1}{2}$. Furthermore, for any buyer i and seller j, (i) the allocation $f_{ij}$ is increasing in buyer i's demand $\alpha_i$ and seller j's supply $\beta_j$, and (ii) the total resource amount $\sum_{j\in\mathcal{S}(i)} f_{ij}$ assigned to buyer i is increasing in her value $v_i$ and $\sum_{i\in\mathcal{B}(j)} f_{ij}$ assigned to seller j is decreasing in her cost $c_j$.*

We outline the proof idea for approximation ratio of $\frac{1}{2}$ in the single-unit version case of problem $\mathcal{P}_1$. We first show that our distributed algorithm converges to the same allocation as a centralized greedy algorithm for solving problem $\mathcal{P}_1$. The centralized algorithm (not optimal) adds every time an edge $(i, j)$ with the maximum weight $w_{ij}$ to the current matching. Such a greedy assignment may affect at most two edges (incident to buyer $i$ or seller $j$) of smaller weights in the optimal allocation. Thus adding each edge $(i, j)$ with weight $w_{ij}$ results in a gap of at most $2w_{ij}$ in the total weight objective as compared to the optimum, and the approximation ratio is $\frac{1}{2}$. The proof for approximation ratio of $\frac{1}{2}$ in the multi-unit case follows by the same argument as in the single-unit case. The detailed proof of Proposition 1 is given in Appendix A.

Besides the worst-case analysis, our extensive numerical analysis shows that the average performance achieved by Algorithm 1 is near-optimal. Consider a D2D market with users uniformly distributed in a circular ground cell with radius of $R = 1$ km. The number of users $M+N$ follows the Poisson point process with mean $\rho = 4000$ and each user chooses to be a buyer or a seller with equal probability. Only when the distance between a buyer and a seller is less than the short communication range $L$, they can connect with each other for resource trading. The buyers' values per unit resource and sellers' costs per unit resource are uniformly distributed over a value set $\mathcal{V} = \{5, 6, \cdots, 10\}$ and a cost set $C = \{0, 1, \cdots, 5\}$, respectively. The demands and supplies respectively are chosen from the set $\Omega = \{1, 2, 3, 4\}$ with equal probabilities.

In Fig. 3, we illustrate the average efficiency of Algorithm 1 compared to the optimum as a function of the communication range $L$. We observe that our Algorithm 1 always achieves average efficiency above 94% of the optimum though its running time is less than 1% of that of the optimal algorithm. The average efficiency first decreases and then increases in the communication range $L$. Intuitively, when L is small (e.g., less than 20 m in Fig. 3), users are sparsely connected, and both Algorithm 1 and the optimum try to match as many pairs as possible if any exists, resulting in little difference in between (or high average efficiency of Algorithm 1). When L is large (e.g., larger than 100 m in Fig. 3), each user has many neighbors and choosing the second best matching is also good. Thus, our Algorithm 1 performs very well and improves as L increases. Similar to Fig. 3, we can also show that the average efficiency of Algorithm 1 first decreases and then increases with the user density, given by $\rho/\pi R^2$, but we skip here due to the page limit.

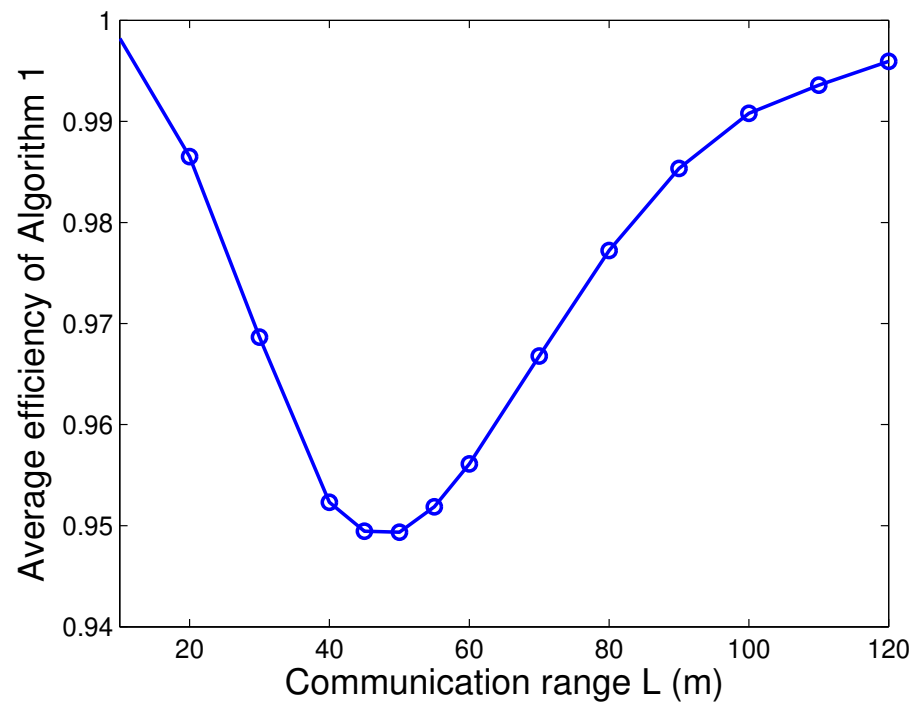

**Figure 3: The average efficiency of Algorithm 1 compared to the optimum as a function of the short communication range $L$ for considering nearby users to communicate.**

Note that users' mutual connectivity improves when we increase the user density or the communication range.

# 4 DISTRIBUTED PRICING MECHANISM

In the last section, Algorithm 1 assumes knowledge of the parameters $\{\alpha_i\}$, $\{v_i\}$, $\{\beta_j\}$ and $\{c_j\}$ to solve problem $\mathcal{P}_1$. But these are private information and may not be reported truthfully unless buyers and sellers are provided with the right pricing incentives. In this section, we first model the utility functions of buyers and sellers and then analyze the selfish user behavior under the traditionally defined prices in double auctions (our 'basic' prices below). Further, we propose a price correction that adjusts the basic prices to make them incentive compatible.

## 4.1 Utility Functions

Remember that we use $p_{ij}^B$ ($p_{ij}^S$) to denote the buying (selling) price for buyer $i$ (seller $j$) between the matched pair $(i, j)$, and assume linear value and cost functions for them. We note that if buyer $i$ is over demanding ($\hat{\alpha}_i > \alpha_i$), she might be assigned with more than what she needs, i.e., $\sum_{j\in\mathcal{S}(i)} f_{ij} > \alpha_i$ without extra benefit but with higher payment. Thus, her utility can be defined as:

$$U_i^B(\alpha_i, v_i) = v_i \min\{\alpha_i, \sum_{j\in\mathcal{S}(i)} f_{ij}\} - \sum_{j\in\mathcal{S}(i)} f_{ij} p_{ij}^B .$$

Similarly, assuming seller $j$ does not over-report her supply ($\hat{\beta}_j > \beta_j$) due to high penalty in case of not fulfilling her commitment, her utility can be expressed as:

$$U_j^S(\beta_j, c_j) = \sum_{i\in\mathcal{B}(j)} f_{ij}(p_{ij}^S - c_j).$$

Observe that a user's utility depends on not only her individual information but also all users' declarations as well as the bipartite matching graph structure $G = (M, N, \{\mathcal{S}(i)\}, \{\mathcal{B}(j)\})$ through the allocations $\{f_{ij}\}$ and prices $\{p_{ij}^B\}$, $\{p_{ij}^S\}$ decided by the mechanism.

## 4.2 Basic Prices

We define the 'basic' price for a transaction between buyer $i$ and seller $j$ as the mean of buyer $i$'s reported value and seller $j$'s reported cost, i.e., $p^B_{ij} = p^S_{ij} = (\hat{v}_i + \hat{c}_j)/2$. Using these prices in our mechanism would be very convenient since (i) prices are derived using only local information, (ii) it distributes surplus fairly among buyers and sellers, and (iii) it is budget balanced. But unfortunately, basic prices are not incentive compatible since buyers (sellers) would like to declare lower values (higher costs) in order to increase their profit. Nevertheless, they play a significant role in our mechanism design.

Assuming that the system platform solves problem $\mathcal{P}_1$ (e.g., by using Algorithm 1) depending on the users' declarations and uses basic prices to compensate buyers and sellers. We define next some key performance measures needed in our incentive compatible design. First note that each user only knows her own information, the distribution of the private information of the other users and the distribution of the random bipartite matching graph, resulting in a *Bayesian game* with incomplete information. Users' private information is assumed to follow the same distribution in general as we consider large-scale markets. By taking expectations over these distributions, we obtain buyer $i$'s expected utility $\bar{U}^B(\alpha_i, v_i, \hat{\alpha}_i, \hat{v}_i)$ assuming that all other users are truthful. This is a function of her private information $(\alpha_i, v_i)$ and declaration $(\hat{\alpha}_i, \hat{v}_i)$. Similarly, the expected utility of seller $j$ is $\bar{U}^S(\beta_j, c_j, \hat{\beta}_j, \hat{c}_j)$. A selfish user's best response is to submit a declaration that maximizes her expected utility. A direct consequence of these utility functions is that buyers are incentivized to report lower values in order to reduce the resulting basic price $(\hat{v}_i + \hat{c}_j)/2$ of the transaction. Similarly, sellers will report higher costs. Hence the above expected utilities are well defined but are not observed in the equilibrium of the game.

Let $\bar{Q}^B(\hat{\alpha}_i, \hat{v}_i)$ ($\bar{Q}^S(\hat{\beta}_j, \hat{c}_j)$) be the expected total resource amount buyer $i$ (seller $j$) obtains in the above Bayesian game assuming all other users being truthful. The above definitions of functions $\bar{U}^B$, $\bar{U}^S$, $\bar{Q}^B$, $\bar{Q}^S$ will be used in our incentive system design that comes next.

## 4.3 Price Correction Scheme

To achieve incentive compatibility, we correct the basic price by adding an incentive price component that only depends on the individual user's declaration to be computed in a distributed manner. In particular, the platform subsidizes buyer $i$ with $g(\hat{\alpha}_i, \hat{v}_i) \geq 0$ per unit resource to buy, leading to a *final* buying price:

$$p^B_{ij} = (\hat{v}_i + \hat{c}_j)/2 - g(\hat{\alpha}_i, \hat{v}_i). \tag{3}$$

Similarly, the platform subsidizes seller $j$ with $h(\hat{\beta}_j, \hat{c}_j) \geq 0$ per unit resource to sell, leading to a *final* selling price:

$$p^S_{ij} = (\hat{v}_i + \hat{c}_j)/2 + h(\hat{\beta}_j, \hat{c}_j). \tag{4}$$

After applying the correction price components $g(\hat{\alpha}_i, \hat{v}_i)$ and $h(\hat{\beta}_j, \hat{c}_j)$, the expected utilities of buyer $i$ and seller $j$ in the Bayesian game become $\bar{U}^B(\alpha_i, v_i, \hat{\alpha}_i, \hat{v}_i) + \bar{g}(\hat{\alpha}_i, \hat{v}_i)$ and $\bar{U}^S(\beta_j, c_j, \hat{\beta}_j, \hat{c}_j) + \bar{h}(\hat{\beta}_j, \hat{c}_j)$, respectively, where $\bar{g}(\hat{\alpha}_i, \hat{v}_i) = g(\hat{\alpha}_i, \hat{v}_i)\bar{Q}^B(\hat{\alpha}_i, \hat{v}_i)$ and $\bar{h}(\hat{\beta}_j, \hat{c}_j) = h(\hat{\beta}_j, \hat{c}_j)\bar{Q}^S(\hat{\beta}_j, \hat{c}_j)$. Our challenge is to find *correction payments* $\bar{g}(\hat{\alpha}_i, \hat{v}_i)$ and $\bar{h}(\hat{\beta}_j, \hat{c}_j)$ that ensure incentive compatibility in the Bayesian game as defined below.

*Definition 1 (Incentive compatibility):* Truthful reporting is a Bayesian Nash equilibrium if no buyer $i$ or seller $j$ can improve her expected utility by unilaterally deviating from the truthful reporting, i.e.,

$$\bar{U}^B(\alpha_i, v_i, \alpha_i, v_i) + \bar{g}(\alpha_i, v_i) \geq \bar{U}^B(\alpha_i, v_i, \hat{\alpha}_i, \hat{v}_i) + \bar{g}(\hat{\alpha}_i, \hat{v}_i), \forall \alpha_i, \hat{\alpha}_i, v_i, \hat{v}_i, \tag{5a}$$

$$\bar{U}^S(\beta_j, c_j, \beta_j, c_j) + \bar{h}(\beta_j, c_j) \geq \bar{U}^S(\beta_j, c_j, \hat{\beta}_j, \hat{c}_j) + \bar{h}(\hat{\beta}_j, \hat{c}_j), \forall \beta_j, \hat{\beta}_j, c_j, \hat{c}_j. \tag{5b}$$

# 5 DESIGN OF CORRECTION PAYMENT

In this section, we design the correction payments $\bar{g}(\hat{\alpha}_i, \hat{v}_i)$ and $\bar{h}(\hat{\beta}_j, \hat{c}_j)$ to satisfy the incentive compatibility constraints in (5). Then, we show that our pricing scheme also satisfies the other three properties (E2)-(E4) as listed at the end of Section 2.2.

## 5.1 Incentive Compatibility Design

In this subsection, we first derive $\bar{g}(\hat{\alpha}_i, \hat{v}_i)$ and $\bar{h}(\hat{\beta}_j, \hat{c}_j)$ assuming that users reveal truthfully their demands or supplies, i.e., $\hat{\alpha}_i = \alpha_i, \forall i \in \mathcal{M}$, $\hat{\beta}_j = \beta_j, \forall j \in \mathcal{N}$. Later we will prove that using these correction payments users have no incentive to misreport on their demands or supplies. For each possible value of demand $\alpha_i$ or supply $\beta_j$, (5) simplifies to:

$$\bar{U}^B(v_i, v_i) + \bar{g}(v_i) \geq \bar{U}^B(v_i, \hat{v}_i) + \bar{g}(\hat{v}_i), \forall v_i, \hat{v}_i, \tag{6a}$$

$$\bar{U}^S(c_j, c_j) + \bar{h}(c_j) \geq \bar{U}^S(c_j, \hat{c}_j) + \bar{h}(\hat{c}_j), \forall c_j, \hat{c}_j. \tag{6b}$$

The following lemma derives from the linearity of the utility functions and is used to further simplify (6).

Lemma 1. *The expected utility functions $\bar{U}^B(v_i, \hat{v}_i)$ for buyer i and $\bar{U}^S(c_j, \hat{c}_j)$ for seller j follow arithmetic progression according to their true value $v_i$ and cost $c_j$, respectively. That is,*

$$\bar{U}^B(v_i, \hat{v}_i) - \bar{U}^B(v_i - 1, \hat{v}_i) = \bar{Q}^B(\hat{v}_i), \tag{7a}$$

$$\bar{U}^S(c_j, \hat{c}_j) - \bar{U}^S(c_j - 1, \hat{c}_j) = -\bar{Q}^S(\hat{c}_j). \tag{7b}$$

The proof is given in Appendix B. By using Proposition 1's monotonicity property of the total resource amount on $\bar{Q}^B(\hat{v}_i)$ and $\bar{Q}^S(\hat{c}_j)$ in (7), we successfully simplify (6) in the following.

Proposition 2. *The incentive compatibility constraints in (6) are equivalent to the following* adjacent *incentive compatibility (AIC) constraints:*

$$\bar{U}^B(v_i, v_i) + \bar{g}(v_i) \geq \bar{U}^B(v_i, \hat{v}_i) + \bar{g}(\hat{v}_i), \forall v_i, \hat{v}_i = v_i - 1, v_i + 1, \tag{8a}$$

$$\bar{U}^S(c_j, c_j) + \bar{h}(c_j) \geq \bar{U}^S(c_j, \hat{c}_j) + \bar{h}(\hat{c}_j), \forall c_j, \hat{c}_j = c_j - 1, c_j + 1. \tag{8b}$$

The proof is given in Appendix C. This proposition states that to induce incentive compatibility over all possible declarations it is enough to guarantee incentive compatibility for declarations adjacent to the true value or cost.[4]

We now propose an iterative algorithm to compute $\bar{g}(\hat{v}_i)$ to satisfy AIC constraint in (8a), which is described in details in Algorithm 2 below. If (8a) under the current correction payments (initially zero) does not hold, there exists a minimum value $\tau$ such that *(8a) is violated at* $v_i = \tau$ *and it holds for all* $v_i < \tau$ (line 4). This means that buyer $i$ with true value $v_i = \tau$ has a greater expected utility when she submits an adjacent value $\tau - 1$ or $\tau + 1$. We can correct that by properly increasing the correction payment $\bar{g}(\tau)$ which is subsidized to this buyer when she submits $\tau$ truthfully (line 7). However, after applying the new $\bar{g}(\tau)$ (line 8-10), (8a) holds for $v_i{=}\tau$ but may be violated for a buyer with $v_i = \tau - 1$ who may over-report the adjacent value $\tau$ to obtain the increased correction payment $\bar{g}(\tau)$ (line 11-12). If this is the case, we further correct it by increasing $\bar{g}(\tau - 1)$ in a way that we don't violate the previously corrected AIC constraint at $\tau$ (line 13). This is possible due to (7a) and the monotonicity property of $\bar{Q}^B(\hat{v}_i)$. Yet this new $\bar{g}(\tau - 1)$ may affect the AIC constraint at $v_i = \tau - 2$ and we proceed similarly to correct any possible violations at $\tau - 2$, $\tau - 3$, etc., until *(8a) holds for all* $v_i \leq \tau$ (line 11-18). After this, we can iteratively increase $\tau$ one by one whiling updating the correction payment for given $\tau$, and will eventually ensure (8a) holds for any value in $\mathcal{V}$.

**Algorithm 2** Correction payment $\bar{g}(\hat{v}_i)$ for any buyer $i$

1: **Initialization**
2: Set $\bar{g}(\hat{v}_i) = 0, \forall \hat{v}_i \in \mathcal{V}$;
3: Set $\bar{U}^B_c(v_i, \hat{v}_i) = \bar{U}^B(v_i, \hat{v}_i) + \bar{g}(\hat{v}_i), \forall v_i, \hat{v}_i \in \mathcal{V}$ to denote the expected utility after correction for buyer $i$;
4: Set $\tau \in \mathcal{V}$ to be the minimum value of $v_i$ such that the AIC constraint in (8a) is violated;
5: **Repeat**
6:   **if** $\max\{\bar{U}^B_c(\tau, \tau+1), \bar{U}^B_c(\tau, \tau-1)\} > \bar{U}^B_c(\tau, \tau)$ **then**
7:     $\bar{g}(\tau) = \bar{g}(\tau) + \max\{\bar{U}^B_c(\tau, \tau+1), \bar{U}^B_c(\tau, \tau-1)\}$ $-\bar{U}^B_c(\tau, \tau)$;
8:     **for** each $v_i \in \mathcal{V}$ **do**
9:       $\bar{U}^B_c(v_i, \tau) = \bar{U}^B(v_i, \tau) + \bar{g}(\tau)$;
10:     **end for**
11:     Set $v = \tau - 1$
12:     **while** $\bar{U}^B_c(v, v) < \bar{U}^B_c(v, v+1)$ **do**
13:       $\bar{g}(v) = \bar{g}(v) + \bar{U}^B_c(v, v+1) - \bar{U}^B_c(v, v)$;
14:       **for** each $v_i \in \mathcal{V}$ **do**
15:         $\bar{U}^B_c(v_i, v) = \bar{U}^B(v_i, v) + \bar{g}(v)$;
16:       **end for**
17:       $v = v - 1$;
18:     **end while**
19:   **end if**
20:   $\tau = \tau + 1$;
21: **Until** $\tau \notin \mathcal{V}$

By running Algorithm 2 for each possible value of demand $\alpha_i$, we obtain the correction payment $\bar{g}(\hat{\alpha}_i, \hat{v}_i)$ that induce all the buyers to truthfully reveal their values. The case of sellers is similar, and we skip the details here due to page limit. A possible implementation is for the platform to run Algorithm 2 offline by simulating the system to obtain the various inputs of the algorithm (e.g., simulate the system based on the distributions of users' private information and the random bipartite matching graph to obtain the expected utility values $\bar{U}^B(\alpha_i, v_i, \hat{\alpha}_i, \hat{v}_i)$, $\bar{U}^S(\beta_j, c_j, \hat{\beta}_j \hat{c}_j)$). Then aware of the correction price rule returned by Algorithm 2, users simply trade with each other in a fully distributed manner, where the final trading prices in (3)-(4) between each matched pair of users only depend on their own declarations.

An additional property of the correction payments returned by Algorithm 2 is that they induce truthful reporting of users' demands or supplies, implying full incentive compatibility in (E1). The proof can be found in Appendix D.

Proposition 3. *With the correction payments* $\bar{g}(\hat{\alpha}_i,\hat{v}_i)$ *and* $\bar{h}(\hat{\beta}_j,\hat{c}_j)$ *returned by Algorithm 2, all the buyers and sellers have no incentive to misreport their demands and supplies.*

## 5.2 Individually Rationality and Budget Balance Design

As buyer $i$ and seller $j$ are truthful now, the final buying price in (3) is always smaller than buyer $i$'s value $v_i$ and the final selling price in (4) is always greater than seller $j$'s cost $c_j$. Thus, our pricing mechanism satisfies individually rationality in (E2).

Moreover, we observe that the platform needs to pay extra $p^S_{ij} - p^B_{ij} = g(\alpha_i, v_i) + h(\beta_j, c_j)$ per unit resource trading for each matched pair $(i, j)$ due to the correction price component. To keep the budget balanced, we propose that the platform charges users a subscription fee (say monthly), in order to recover the correction payments paid to all users. In the case of symmetric users, this subscription fee should be the same and could be determined by the platform running offline simulations or recording the actual correction payments. If the users belong to different classes, the platform can also determine a fair subscription fee. Moreover, although users pay back in the form of the subscription fee the total correction payments they receive over time, they are still left with positive surplus since the basic price component is always less than the actual value for the buyers and higher than the cost for the sellers. Thus, properties (E3) and (E4) are both satisfied.

# 6 EXTENSION TO MULTIPLE TRADING ROUNDS

In practice, our mechanism runs repetitively over time with users' random arrival, movement and departure. Some interesting questions are whether connections between the same pair of users may persist over multiple rounds and how frequently to do trading.

## 6.1 Low Switching Cost for Allocation in Algorithm 1

Consider two subsequent rounds of the trading algorithm, where in round 1 buyer $i$ is matched with seller $j$, and both users participate in round 2. Because new users join the system or existing users depart, the allocation algorithm might choose different pairings for buyer $i$ and seller $j$ in round 2, causing 'switching cost' (extra coordination messages, establishment of new direct connections, etc.). Our simulations in Fig. 4 show that Algorithm 1 performs significantly better than the optimal algorithm in term of switching cost defined similarly to [9] as *the mean number of new matched pairs between any two subsequent rounds*.

[4] In our model, we assume value set $\mathcal{V}$ and cost set $C$ contain consecutive integers and 1 is the minimum gap size. Thus, the adjacent values for $v_i$ are $v_i - 1$ and $v_i + 1$, and the adjacent costs for $c_j$ are $c_j - 1$ and $c_j + 1$.

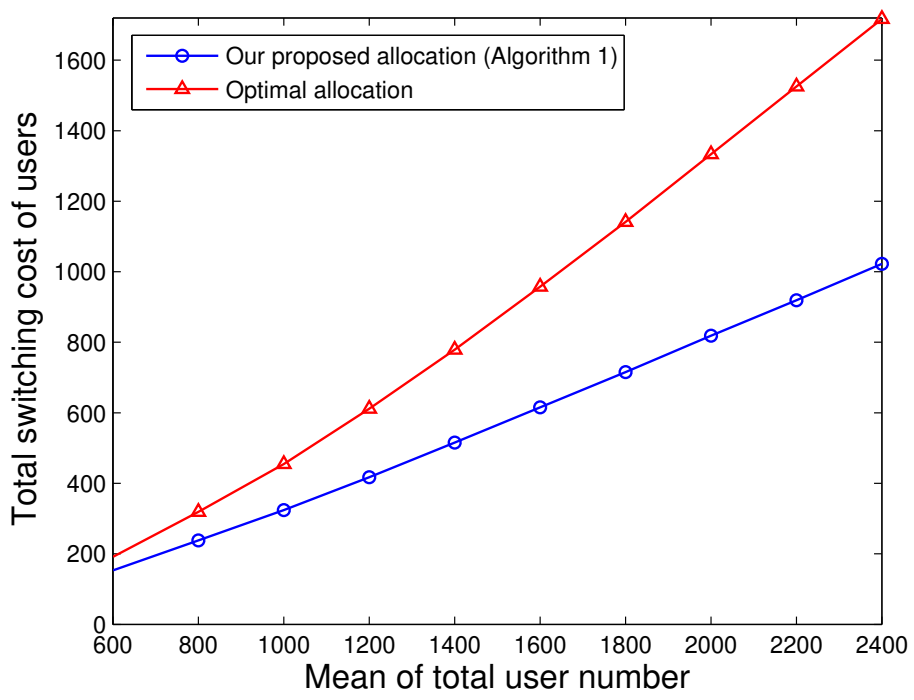

**Figure 4: The total switching costs under our proposed Algorithm 1 and the optimal algorithm versus the mean of total user number $\rho$. Here we set $R = 1$ km and $L = 100$ m.**

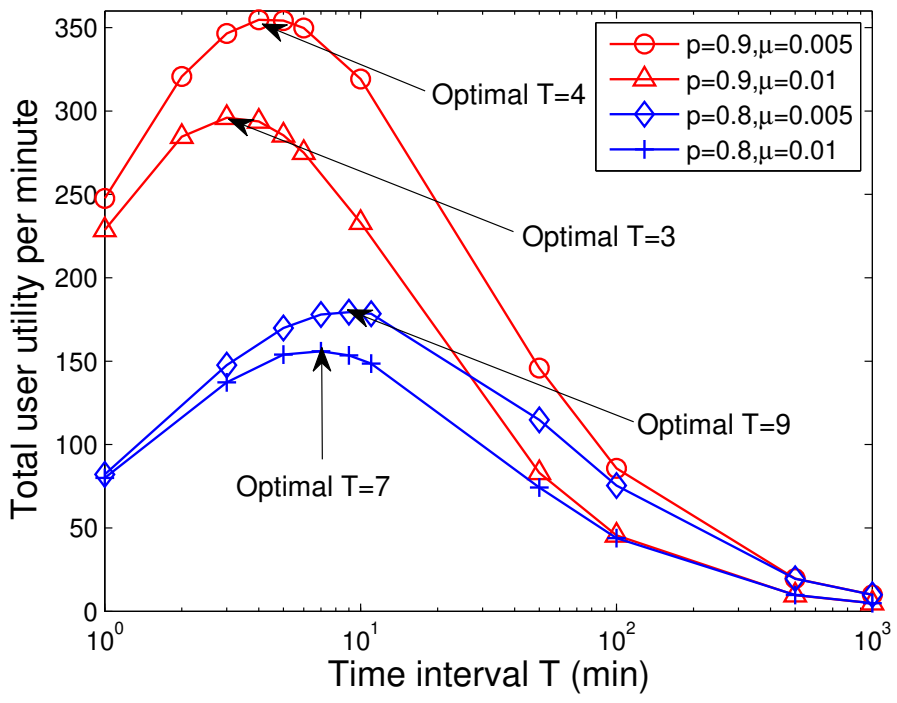

**Figure 5: The time-average total user utility versus the trading time interval $T$ for different values of probability to trade again $p$ and departure rate $\mu$. Here we set $\lambda = 20$ per minute, $R = 1$ km and $L = 100$ m.**

In our simulations, we consider the same market model as in Section 3 that all users are uniformly distributed in a circular ground cell with radius of $R = 1$ km. The total number of users follows the Poisson point process with mean $\rho$ and the short communication range $L = 100$ m between users. Moreover, we assume that users participating in round 1 leave with probability 0.2 before the beginning of round 2, and new users of types chosen uniformly at random join at rate $0.2\rho$ so that the average number of users remains the same. In Fig. 4, we illustrate the total switching costs against the mean of total user number $\rho$ under our Algorithm 1 and the optimal algorithm. Observe that our algorithm has lower switching cost as compared to the optimal algorithm especially when the network size is large, leading to more than 40% cost savings when there are more than 2000 users. This happens because Algorithm 1 keeps the locally optimal neighbors for each user, and does not propagate easily small changes across the network especially when the network size is large. This is another advantage of our Algorithm 1 besides its near-optimality and low-complexity.

## 6.2 Optimal Trading Frequency

How should the platform choose the length $T$ (e.g. in minutes) of the time interval between consecutive rounds? If $T$ is small, few new participants will join the system and many of the existing users may have no new resources to trade as trade takes place so frequently. If $T$ is large, new users may join, but trading will be infrequent and impatient users may leave. To understand this tradeoff, we consider a simple scenario that a fixed number $\lambda$ of new users arrive per minutes, each user being a buyer or a seller with equal probability and trading a single resource unit. A user that participates will leave the system after an exponential time with departure rate $\mu$. Further, for those who stayed since last trading, they are still interested to trade again in the new round with probability $p < 1$ and thus an average user has resources to trade for $1/(1-p)$ rounds even if her participation time is long enough. In this system, if $K$ denotes the average number of participants in the steady-state, we have $K = Kpe^{-\mu T} + \sum_{t=0}^{T-1} \lambda e^{-\mu(T-t)}$, where the first term on the right-hand-side of the equation tells how many out of $K$ participators in last trading are still interested in new trading (with probability $p$) and haven't left (with probability $e^{-\mu T}$), and the second term tells how many new users joining after last trading will participate in the new trading. Thus, $K = \frac{\lambda e^{-\mu}(1-e^{-\mu T})}{(1-pe^{-\mu T})(1-e^{-\mu})}$ and we can approximate the time-average total user utility as $\mathbb{E}_{M,N\sim B(K,0.5)} \sum_{i\in\mathcal{M}} \sum_{j\in\mathcal{S}(i)} w_{ij} f_{ij}/T$.

In Fig. 5, we show the time-averaged total user utility versus $T$ for different values of $p$ and $\mu$. It first increases and then decreases with $T$ for all the four parameter settings, telling the tradeoff between trading opportunity (for a stably good number of users to encounter and trade) and the waiting cost (for impatient user departures in the meantime). Moreover, the optimal $T$ decreases with $p$, since a greater $p$ keeps more users from previous rounds in the system and a smaller $T$ is enough to ensure enough users for trading in the new round. It decreases with $\mu$ as users leave the market more frequently and we need a smaller $T$ to keep them in trading.

# 7 CONCLUSIONS

In this paper, we develop a novel distributed double auction mechanism to enable direct resource trading in a large-scale D2D trading market. This mechanism consists of a resource allocation algorithm to match demand and supply and a pricing mechanism to determine the final trading prices. We first propose an efficient distributed allocation algorithm to solve the resource allocation problem with D2D assignment constraints. Then, we design a distributed pricing mechanism to motivate users to report their private information truthfully which is also individually rational and ex-ante budget balanced. Finally, we extend the proposed mechanism to multiple rounds and examine the switching cost between two subsequent rounds as well as the best trading frequency. Our proposed allocation algorithm, besides being nearly optimal has advantages in terminating in a small number of steps and of reducing switching cost.

The goal of the paper was to address theoretical issues in the design of the market, and thus many details on implementation are left out. These include how to implement the central subscription system, the fee payments, the case of asymmetric users, verifiability of the actual delivery of the goods from sellers to buyers, etc. Another direction for further research is using an adaptive learning algorithm to converge to the solution of (5) for computing the incentive compatible payments instead of using simulations.

## REFERENCES

[1] Panayotis Antoniadis, Costas Courcoubetis, and Robin Mason. 2004. Comparing economic incentives in peer-to-peer networks. *Computer Networks* 46, 1 (2004), 133 – 146. https://doi.org/10.1016/j.comnet.2004.03.021 Internet Economics: Pricing and Policies.

[2] Kenneth J Arrow. 1979. The property rights doctrine and demand revelation under incomplete information. In *Economics and human welfare*. Elsevier, 23–39.

[3] Dimitri P Bertsekas and David A Castanon. 1989. The auction algorithm for the transportation problem. *Annals of Operations Research* 20, 1 (1989), 67–96.

[4] Suzhi Bi, Yong Zeng, and Rui Zhang. 2016. Wireless powered communication networks: an overview. *IEEE Wireless Communications* 23, 2 (April 2016), 10–18. https://doi.org/10.1109/MWC.2016.7462480

[5] Yinghao Guo, Lingjie Duan, and Rui Zhang. 2017. Cooperative Local Caching Under Heterogeneous File Preferences. *IEEE Transactions on Communications* 65, 1 (Jan 2017), 444–457. https://doi.org/10.1109/TCOMM.2016.2620164

[6] Jaap-Henk Hoepman. 2004. Simple distributed weighted matchings. *arXiv preprint cs/0410047* (2004).

[7] George Iosifidis, Lin Gao, Jianwei Huang, and Leandros Tassiulas. 2015. A Double-Auction Mechanism for Mobile Data-Offloading Markets. *IEEE/ACM Transactions on Networking* 23, 5 (Oct 2015), 1634–1647. https://doi.org/10.1109/TNET.2014.2345875

[8] Amin M Khan, Xavier Vilaça, Luís Rodrigues, and Felix Freitag. 2016. A Distributed Auctioneer for Resource Allocation in Decentralized Systems. In *2016 IEEE 36th International Conference on Distributed Computing Systems (ICDCS)*. 201–210. https://doi.org/10.1109/ICDCS.2016.38

[9] Paul Klemperer. 1987. The competitiveness of markets with switching costs. *The RAND Journal of Economics* (1987), 138–150.

[10] Yunpeng Li, Costas Courcoubetis, and Lingjie Duan. 2017. Dynamic Routing for Social Information Sharing. *IEEE Journal on Selected Areas in Communications* 35, 3 (2017), 571–585.

[11] Yunpeng Li, Costas Courcoubetis, and Lingjie Duan. 2019. Recommending Paths: Follow or Not Follow?. In *IEEE INFOCOM 2019 - IEEE Conference on Computer Communications*. 928–936.

[12] R.Preston McAfee. 1992. A dominant strategy double auction. *Journal of Economic Theory* 56, 2 (1992), 434 – 450. https://doi.org/10.1016/0022-0531(92)90091-U

[13] Noam Nisan and Amir Ronen. 2001. Algorithmic mechanism design. *Games and Economic behavior* 35, 1-2 (2001), 166–196.

[14] Robert Preis. 1999. Linear time 1/2-approximation algorithm for maximum weighted matching in general graphs. In *Annual Symposium on Theoretical Aspects of Computer Science*. Springer, 259–269.

[15] Lingjun Pu, Xu Chen, Jingdong Xu, and Xiaoming Fu. 2016. D2D Fogging: An Energy-Efficient and Incentive-Aware Task Offloading Framework via Network-assisted D2D Collaboration. *IEEE Journal on Selected Areas in Communications* 34, 12 (Dec 2016), 3887–3901. https://doi.org/10.1109/JSAC.2016.2624118

[16] Ming Tang, Shou Wang, Lin Gao, Jianwei Huang, and Lifeng Sun. 2017. MOMD: A multi-object multi-dimensional auction for crowdsourced mobile video streaming. In *IEEE INFOCOM 2017 - IEEE Conference on Computer Communications*. 1–9. https://doi.org/10.1109/INFOCOM.2017.8057025

[17] Xuehe Wang, Lingjie Duan, and Rui Zhang. 2016. User-Initiated Data Plan Trading via a Personal Hotspot Market. *IEEE Transactions on Wireless Communications* 15, 11 (Nov 2016), 7885–7898. https://doi.org/10.1109/TWC.2016.2608957

[18] Yang Zhang, Chonho Lee, Dusit Niyato, and Ping Wang. 2013. Auction Approaches for Resource Allocation in Wireless Systems: A Survey. *IEEE Communications Surveys Tutorials* 15, 3 (Third 2013), 1020–1041. https://doi.org/10.1109/SURV.2012.110112.00125

[19] Liang Zheng, Carlee Joe-Wong, Chee Wei Tan, Sangtae Ha, and Mung Chiangs. 2015. Secondary markets for mobile data: Feasibility and benefits of traded data plans. In *2015 IEEE Conference on Computer Communications (INFOCOM)*. 1580–1588. https://doi.org/10.1109/INFOCOM.2015.7218537

# A PROOF OF PROPOSITION 1

## A.1 Approximation Ratio of $\frac{1}{2}$

First, for our multi-unit allocation problem $\mathcal{P}_1$, we consider a centralized greedy algorithm that includes the edge with maximum weight sequentially as in the single-unit greedy matching algorithm. In the centralized algorithm, every time we add an edge $(i, j)$ with the current maximum weight, we update allocation $f_{ij} = \min\{\alpha_i, \beta_j\}$ units of resources, which is upper bounded by both buyer $i$'s demand and seller $j$'s supply. Then, buyer $i$ updates her unsatisfied demand as $\alpha_i = \alpha_i - \min\{\alpha_i, \beta_j\}$ and seller $j$ updates her available supply as $\beta_j = \beta_j - \min\{\alpha_i, \beta_j\}$. If the buyer or the seller is fully satisfied (i.e. $\alpha_i = 0$ or $\beta_j = 0$), she and all her edges will be removed from the graph.

Next, we prove that Algorithm 1 can converge to the same allocation as the prior centralized greedy algorithm. In the first iteration of Algorithm 1, buyer $i$ first sorts her neighboring sellers in a non-increasing weight order, denoted as $\{j_1, j_2, \cdots, j_{\min\{\alpha_i, |\mathcal{S}(i)|\}}\}$ with $w_{ij_1} \geq w_{ij_2} \geq \cdots \geq w_{ij_{\min\{\alpha_i, |\mathcal{S}(i)|\}}}$. Then she computes the maximum allocation can be assigned to seller $j_1$, given by $\max\{\alpha_i, \beta_{j_1}\}$, and sends the corresponding request to seller $j_1$. If an edge $(i, j)$ has the local maximum weight from both sides of buyer $i$ and seller $j$ (i.e., seller $j$ is the best seller $j_1$ of buyer $i$ and buyer $i$ is the best buyer $i_1$ of seller $j$), it must be added with $\max\{\alpha_i, \beta_j\}$ units of resource in Algorithm 1. Moreover, this edge also must be added with $\max\{\alpha_i, \beta_j\}$ units of resource in the centralized algorithm. This is because the edges added before $(i, j)$ cannot be incident to $i$ or $j$ as $(i, j)$ has the local maximum weight from both sides and thus buyer $i$'s unsatisfied demand and seller $j$'s available supply do not change until adding $(i, j)$.

Except the edges with local maximum weights, Algorithm 1 may also add some second best edges in the first iteration. For example, if buyer $i$'s demand $\alpha_i$ is higher than her best seller $j_1$'s supply $\beta_{j_1}$, she will ask seller $j_2$ for $f^B_{ij_2} = \min\{\beta_{j_2}, \alpha_i - \beta_{j_1}\}$ units of resources to satisfy her remaining demand. Note that in the centralized algorithm, the only edge incident to buyer $i$ that can be added before edge $(i, j_2)$ is edge $(i, j_1)$ and the allocation to edge $(i, j_1)$ is bounded by $\max\{\alpha_i, \beta_{j_1}\} = \beta_{j_1}$. Thus, before adding $w_{ij_2}$ in the centralized algorithm, buyer $i$'s unsatisfied demand is not less than $\alpha_i - \beta_{j_1}\}$. From the side of buyer $i$, her request $f^B_{ij_2} = \min\{\beta_{j_2}, \alpha_i - \beta_{j_1}\}$ to seller $j_2$ in Algorithm 1 is not larger than her available demand when adding $(i, j_2)$ in the centralized algorithm. Similarly, from the side of seller $j_2$, her request $f^S_{ij_2}$ to buyer $i$ in Algorithm 1 is also not larger than her available supply when adding $(i, j_2)$ in the centralized algorithm. Thus, at least $\min\{f^B_{ij_2}, f^S_{ij_2}\}$ units of resource must be added for edge $(i, j_2)$ in the centralized algorithm.

We can conclude that the allocation obtained in the first iteration of Algorithm 1 is not larger than the final allocation returned by the centralized algorithm. Then, by mathematical induction, we prove this holds for each iteration of Algorithm 1. Moreover, Algorithm 1 and the centralized algorithm have the same convergence condition that all users are fully satisfied or see no available neighbor. Thus, Algorithm 1 will not converge until it obtains the same allocation as the centralized algorithm.

Finally, we prove that the centralized greedy algorithm can achieve an approximation ratio of $\frac{1}{2}$ using the same proof idea in the single-unit case. The centralized algorithm adds every time an edge $(i, j)$ with the maximum weight $w_{ij}$ for $f_{ij}$ units of resources. As compared to the optimum, buyer $i$ loses at most $f_{ij}$ units of allocation with other neighboring sellers due to her demand is reduced by $f_{ij}$, and seller $j$ loses at most $f_{ij}$ allocation with other neighboring buyers due to her supply is reduced by $f_{ij}$. Thus, such a greedy assignment may affect at most $2f_{ij}$ units of allocation (incident to buyer $i$ or seller $j$) with smaller weights than $w_{ij}$. Thus adding each edge $(i, j)$ for $f_{ij}$ units of resources results in a gap of at most $2w_{ij}f_{ij}$ in the total weight objective as compared to the optimum, and the approximation ratio is $\frac{1}{2}$.

## A.2 $f_{ij}$ Increasing in Demand $\alpha_i$ and Supply $\beta_j$

As we discussed in Appendix A.1, Algorithm 1 converges to the same allocation as the centralized greedy algorithm given. Therefore, we turn to prove that the allocation $f_{ij}$ is increasing in buyer $i$'s demand $\alpha_i$ and seller $j$'s supply $\beta_j$ in Algorithm 1.

Given that buyer $i$ submits a higher demand $\alpha_i'$ than the original $\alpha_i$, we discuss the new allocation $f_{ij}'$ returned by the centralized greedy algorithm under the following two cases. These are defined depending on whether the original allocation $\sum_{j\in\mathcal{S}(i)} f_{ij}$ partially or fully satisfies the demand $\alpha_i$. In the former case, all the allocations corresponding to buyer $i$ are constrained by the available supplies of her matched sellers, not by her unsatisfied demand. Submitting a higher demand $\alpha_i'$ than $\alpha_i$ has no effect on the returned allocation. In the latter case, let $j_{min}$ denote the seller with the minimum weight among the matched sellers of buyer $i$ such that $w_{ij_{min}} = \min\{w_{ij} : f_{ij} > 0\}$. For seller $j \neq j_{min}$ with $f_{ij} > 0$, the allocation $f_{ij}$ is also constrained by seller $j$'s available supply and thus the new $f_{ij}' = f_{ij}$. For seller $j = j_{min}$, $f_{ij}$ is constrained either by seller $j$'s available supply or buyer $i$'s unsatisfied demand. Since seller $j$'s available supply stays the same and buyer $i$'s unsatisfied demand is increased, we have $f_{ij_{min}}' \geq f_{ij_{min}}$. For seller $j$ with $f_{ij} = 0$, it is obvious that $f_{ij}' \geq f_{ij} = 0$.

Thus, we can conclude that the returned allocation satisfies $f_{ij}' \geq f_{ij}, \forall j \in \mathcal{S}(i)$ when $\alpha_i' > \alpha_i$. The proof for sellers follows by similar arguments.

## A.3 $\sum_{j\in\mathcal{S}(i)} f_{ij}$ Increasing in Value $v_i$ and $\sum_{i\in\mathcal{B}(j)} f_{ij}$ Decreasing in Cost $c_j$.

If buyer $i$ submits a higher value $v_i'$ than the original $v_i$, the weights of the edges incident to buyer node $i$ are all increased by $v_i' - v_i$. Thus, it is sufficient for us to prove that the total resource amount $\sum_{j\in\mathcal{S}(i)} f_{ij}$ increases as weight $w_{ij}$ increases for any seller $j \in \mathcal{S}(i)$. Moreover, we consider the centralized greedy algorithm instead of Algorithm 1 as in Appendix A.2.

In the centralized greedy algorithm, when weight $w_{ij}$ of edge $(i, j)$ increases, the order of this edge to be added is moved up or at least stays the same. Thus, we have the new allocation $f_{ij}' \geq f_{ij}$. If $f_{ij}' = f_{ij}$, adding edge $(i, j)$ in advance has no influence on the following adding of the remaining edges since adding it according to the original order leads to the same amount of reduction for buyer $i$'s demand seller $j$'s supply. If $f_{ij}' > f_{ij}$, the additional $f_{ij}' - f_{ij}$ allocation may influence the final returned allocation.

First, we consider the case that $f_{ij}' = f_{ij} + 1$. In this case, both buyer $i$'s unsatisfied demand and seller $j$'s available supply are decreased by 1. The one unit reduction of seller $j$'s supply causes a chain effect represented by $\{(j, i_1), (i_1, j_1), (j_1, i_2) \cdots\}$ along which a single unit flow denotes the different allocation between $\{f_{ij}'\}$ and $\{f_{ij}\}$. Note that in this chain, allocation to an edge from buyer to seller (e.g. $(i_1, j_1)$) is increased by 1 and the allocation to an edge from seller to buyer (e.g. $(j, i_1)$) is decreased by 1. For each user node in this chain, only the start node's and the end node's total allocation amounts may change by 1. Even if the chain ends at the node $i$, the decreased one unit allocation to buyer $i$ can be cancelled by the additional one unit allocation to edge $(i, j)$. The total resource amount assigned to buyer $i$ does not decrease.

Similarly, the one unit reduction of buyer $i$'s unsatisfied demand causes a chain effect represented as $\{(i, j_1'), (j_1', i_1'), (i_1', j_2') \cdots\}$. But, in this chain, the allocation to an edge from buyer to seller (e.g. $(i, j_1')$) is decreased by 1 and the allocation to an edge from seller to buyer (e.g. $(j_1', i_1')$) is increased by 1. Similarly, for each user node in this chain, only the start node's and the end node's total allocation amounts may change by 1. If the chain is empty or ends at the start node $i$, buyer $i$'s total allocation amount does not change. Otherwise, its allocation amount is decreased by 1 due to the allocation to edge $(i, j_1')$ decreased by 1. This happens only when the amount of units assigned to edge $(i, j_1')$ is constrained by buyer $i$'s unsatisfied demand or seller $j_1'$'s available supply is decreased by 1. In the former case, buyer $i$ is fully satisfied after adding $(i, j_1')$ and we have $\sum_{j\in\mathcal{S}(i)} f_{ij}' = \alpha_i \geq \sum_{j\in\mathcal{S}(i)} f_{ij}$ for sure. In the latter case, the only reason for seller $j_1$'s available supply decreased by 1 is that another chain starting from seller $j$ ends at node $j_1$, not at $i$. Thus, the chain starting from seller $j$ does not change the allocation to buyer $i$. The decreased one unit allocation to buyer $i$ caused by the chain starting from buyer $i$ can be cancelled by the additional one unit allocation to edge $(i, j)$ and hence $\sum_{j\in\mathcal{S}(i)} f_{ij}' \geq \sum_{j\in\mathcal{S}(i)} f_{ij}$ still holds.

In sum, we prove $\sum_{j\in\mathcal{S}(i)} f_{ij}' \geq \sum_{j\in\mathcal{S}(i)} f_{ij}$ given $f_{ij}' = f_{ij} + 1$. For the more general case $f_{ij}' > f_{ij}$, we can further prove the total amount of resources allocated to buyer $i$ keep non-decreasing when we increase allocation to edge $(i, j)$ from $f_{ij}$ one by one until reaching $f_{ij}'$ similarly. The proof of the monotonicity property of $\sum_{j\in\mathcal{S}(i)} f_{ij}$ in $\hat{v}_i$ is completed.

The proof for sellers follows by similar arguments.

# B PROOF OF LEMMA 1

Since the mechanism runs depending on the local declarations, we have that if a buyer does not change her declared value $\hat{v}_i$ and demand $\hat{\alpha}_i$, the allocation $\{f_{ij}\}$ and the final buying price $\{p_{ij}^B\}$ for her will not change, either. Moreover, given buyers reveal their demand truthfully, buyer $i$ can only be assigned with no more than what she reports (i.e., $\sum_{j\in\mathcal{S}(i)} f_{ij} \leq \hat{\alpha}_i < \alpha_i$). Therefore, from the utility functions given in Section 4.1, we have that the difference of the average utilities is equal to the difference of the true values times the average total amount of allocated resources

$\bar{Q}^B(\hat{v}_i) = \mathbb{E}[\sum_{j\in\mathcal{S}(i)} f_{ij}]$:

$$\bar{U}^B(v_i, \hat{v}_i) - \bar{U}^B(v_i', \hat{v}_i) = (v_i - v_i')\bar{Q}^B(\hat{v}_i)$$

The proof for sellers follows by similar arguments.

# C PROOF OF PROPOSITION 2

Let $\bar{U}_c^B(v_i, \hat{v}_i) = \bar{U}^B(v_i, \hat{v}_i) + \bar{g}(\hat{v}_i)$ denote the expected utility after correction for any buyer $i$. Then, (6a) and (8a) can be rewritten as follows, respectively:

$$\bar{U}_c^B(v_i, v_i) \geq \bar{U}_c^B(v_i, \hat{v}_i), \quad \forall v_i, \hat{v}_i, \tag{9a}$$

$$\bar{U}_c^B(v_i, v_i) \geq \bar{U}_c^B(v_i, \hat{v}_i), \quad \forall v_i, \hat{v}_i = v_i - 1, v_i + 1. \tag{9b}$$

To prove (6a) and (8a) are equivalent, we only need to show that the above two inequalities are equivalent. First observe that the function $\bar{U}_c^B$ satisfying (9a) must satisfy (9b), which can be easily proved by substituting $\hat{v}_i = v_i - 1, v_i + 1$ into (9a). As for the other direction of equivalence, it requires some more work as shown below.

Suppose that the function $\bar{U}_c^B$ satisfies (9b). The 'upward' part of the AIC constraint (9b) tells that buyer $i$ with true value $v_i$ will not obtain a higher expected utility when she submits an 'upward' adjacent value $v_i + 1$, i.e.,

$$\bar{U}_c^B(v_i, v_i) \geq \bar{U}_c^B(v_i, v_i + 1). \tag{10}$$

Further, by increasing $v_i$ by 1, we also obtain:

$$\bar{U}_c^B(v_i + 1, v_i + 1) \geq \bar{U}_c^B(v_i + 1, v_i + 2). \tag{11}$$

Combining this inequality with (7a), we can further derive that $\bar{U}_c^B(v_i, v_i + 1) \geq \bar{U}_c^B(v_i, v_i + 2)$ as follows:

$$\begin{aligned}\bar{U}_c^B(v_i, v_i + 1) &= \bar{U}_c^B(v_i + 1, v_i + 1) - \bar{Q}^B(v_i + 1) \\ &\geq \bar{U}_c^B(v_i + 1, v_i + 2) - \bar{Q}^B(v_i + 2) = \bar{U}_c^B(v_i, v_i + 2),\end{aligned} \tag{12}$$

where the inequality uses (11) and the monotonicity property of $\bar{Q}^B(\hat{v}_i)$ in $\hat{v}_i$.

Similarly to (10) and (12), we can prove that $\bar{U}_c^B(v_i,v) \geq \bar{U}_c^B(v_i,v + 1)$ for any value $v \geq v_i$. Thus, if $\hat{v}_i > v_i$, we have:

$$\begin{aligned}&\bar{U}_c^B(v_i, v_i) \geq \bar{U}_c^B(v_i, v_i + 1), \\ &\bar{U}_c^B(v_i, v_i + 1) \geq \bar{U}_c^B(v_i, v_i + 2), \\ &\cdots \\ &\bar{U}_c^B(v_i, \hat{v}_i - 1) \geq \bar{U}_c^B(v_i, \hat{v}_i).\end{aligned}$$

By summing these above inequalities, we finally obtain $\bar{U}_c^B(v_i, v_i) \geq \bar{U}_c^B(v_i, \hat{v}_i)$ for $\hat{v}_i > v_i$. Then, we can also prove it for $\hat{v}_i < v_i$ by using the 'downward' adjacent part of (9b) that implies buyer $i$ with true value $v_i$ will not obtain a higher expected utility when she submits an 'downward' adjacent value $v_i - 1$. (9a) now holds and the proof of the other direction of equivalence is completed. Therefore, we can conclude that (9a) is equivalent to (9b) for buyers.

The proof for sellers follows by similar arguments.

# D PROOF OF PROPOSITION 3

As we mention in Section 4.2, buyers would like to declare lower values to reduce their buying prices. If buyer $i$ with value $v_i$ has a greater expected utility when under-reporting the adjacent value $v_i - 1$, Algorithm 2 (line 7) will increase the correction payment $\bar{g}(\alpha_i, v_i)$ to make buyer $i$ obtain the *same expected utility* when she report $v_i$ truthfully. Thus, any non-zero (being increased) $\bar{g}(\alpha_i, v_i)$ returned by Algorithm 2 satisfies:

$$\bar{U}_c^B(\alpha_i, v_i, \alpha_i, v_i) = \bar{U}_c^B(\alpha_i, v_i, \alpha_i, v_i - 1), \tag{13}$$

where $\bar{U}_c^B(\alpha_i, v_i, \hat{\alpha}_i, \hat{v}_i) = \bar{U}^B(\alpha_i, v_i, \hat{\alpha}_i, \hat{v}_i) + \bar{g}(\hat{\alpha}_i, \hat{v}_i)$ denotes the expected utility after correction for any buyer $i$. Using this result, we will prove any buyer $i$ has no incentive to under-report or over-report her demand, respectively.

## D.1 Under-reporting Demand ($\hat{\alpha}_i < \alpha_i$)

In this case, buyer $i$ reports a lower demand than the true one and she can only be assigned with no more than what she reports (i.e., $\sum_{j\in\mathcal{S}(i)} f_{ij} \leq \hat{\alpha}_i < \alpha_i$) even if she has a high true demand. Thus, she obtains exactly the same utility as if her true demand is what she reports, i.e.,

$$\bar{U}_c^B(\alpha_i, v_i, \hat{\alpha}_i, \hat{v}_i) = \bar{U}_c^B(\hat{\alpha}_i, v_i, \hat{\alpha}_i, \hat{v}_i), \forall v_i, \hat{v}_i, \hat{\alpha}_i < \alpha_i. \tag{14}$$

Since the correction payment returned by Algorithm 2 satisfies the incentive compatibility constraints for values when buyers truthfully reveal their demands, we have that $\bar{U}_c^B(\alpha_i, v_i, \hat{\alpha}_i, v_i) \geq \bar{U}_c^B(\alpha_i, v_i, \hat{\alpha}_i, \hat{v}_i)$ for all $\hat{v}_i$ when $\hat{\alpha}_i = \alpha_i$. Combining this inequality with (14), we further derive that:

$$\begin{aligned}&\bar{U}_c^B(\alpha_i, v_i, \hat{\alpha}_i, \hat{v}_i) = \bar{U}_c^B(\hat{\alpha}_i, v_i, \hat{\alpha}_i, \hat{v}_i) \\ &\leq \bar{U}_c^B(\hat{\alpha}_i, v_i, \hat{\alpha}_i, v_i) = \bar{U}_c^B(\alpha_i, v_i, \hat{\alpha}_i, v_i), \forall v_i, \hat{v}_i, \hat{\alpha}_i < \alpha_i.\end{aligned}$$

Thus, it is best for buyers to truthfully reveal their values in the under-reporting case. Without concern for misreporting of values, we can prove that buyers have no incentive to under-report their demands by simply proving:

$$\bar{U}_c^B(\alpha_i, v_i, \alpha_i, v_i) \geq \bar{U}_c^B(\hat{\alpha}_i, v_i, \hat{\alpha}_i, v_i), \forall v_i, \hat{\alpha}_i < \alpha_i. \tag{15}$$

To prove that, we first find $\tau$, the minimum value such that the correction $\bar{g}(\hat{\alpha}_i, \tau)$ returned by Algorithm 2 is not zero. For $v_i < \tau$, the correction $\bar{g}(\hat{\alpha}_i, v_i)$ is zero and hence,

$$\begin{aligned}&\bar{U}_c^B(\alpha_i, v_i, \alpha_i, v_i) \overset{(a)}{\geq} \bar{U}^B(\alpha_i, v_i, \alpha_i, v_i) \\ &\overset{(b)}{\geq} \bar{U}^B(\hat{\alpha}_i, v_i, \hat{\alpha}_i, v_i) \overset{(c)}{=} \bar{U}_c^B(\hat{\alpha}_i, v_i, \hat{\alpha}_i, v_i), \forall \hat{\alpha}_i < \alpha_i,\end{aligned}$$

where (a) is due to non-negative correction and (c) is because $\bar{g}(\hat{\alpha}_i, v_i) = 0$. (b) uses monotonicity property of $\bar{U}^B(\hat{\alpha}_i, v_i, \hat{\alpha}_i, v_i)$ in $\hat{\alpha}_i$, which can be derived by the monotonicity property of $f_{ij}$ in Proposition 1.

Now, given $\bar{U}_c^B(\alpha_i, v_i, \alpha_i, v_i) \geq \bar{U}_c^B(\hat{\alpha}_i, v_i, \hat{\alpha}_i, v_i)$ for all $v_i < \tau$, we further prove $\bar{U}_c^B(\alpha_i, \tau, \alpha_i, \tau) \geq \bar{U}_c^B(\hat{\alpha}_i, \tau, \hat{\alpha}_i, \tau)$ as follows:

$$\begin{aligned}&\bar{U}_c^B(\alpha_i, \tau, \alpha_i, \tau) \overset{(d)}{\geq} \bar{U}_c^B(\alpha_i, \tau, \alpha_i, \tau - 1) \\ &= \bar{U}_c^B(\alpha_i, \tau - 1, \alpha_i, \tau - 1) + \bar{Q}^B(\alpha_i, \tau - 1) \\ &\overset{(e)}{\geq} \bar{U}_c^B(\hat{\alpha}_i, \tau - 1, \hat{\alpha}_i, \tau - 1) + \bar{Q}^B(\hat{\alpha}_i, \tau - 1) \\ &= \bar{U}_c^B(\hat{\alpha}_i, \tau, \hat{\alpha}_i, \tau - 1) \overset{(f)}{=} \bar{U}_c^B(\hat{\alpha}_i, \tau, \hat{\alpha}_i, \tau), \forall \hat{\alpha}_i < \alpha_i,\end{aligned}$$

where (d) uses the AIC of function $\bar{U}_c^B$ when truthfully reporting demand, (e) uses the given condition and the monotonicity property of $\bar{Q}^B(\hat{\alpha}_i, \hat{v}_i)$ in $\hat{\alpha}_i$, (f) is derived by (13) and the other two equalities are based on (7a).

Then, by using mathematical induction, we can finally prove $\bar{U}_c^B(\alpha_i, v_i, \alpha_i, v_i) \geq \bar{U}_c^B(\hat{\alpha}_i, v_i, \hat{\alpha}_i, v_i)$ for all $v_i$ and $\hat{\alpha}_i < \alpha_i$ as in (15). Therefore, we conclude that buyers have no incentive to under-report their demands. The proof of the under-reporting case for buyers is completed.

## D.2 Over-reporting Demand ($\hat{\alpha}_i > \alpha_i$)

First note that, in our system model, if buyer $i$ gets more allocation than her actual demand $\alpha_i$, she will only use $\alpha_i$ resources and the rest is useless. Then, we prove that by submitting a higher demand $\hat{\alpha}_i > \alpha_i$, the expected amount of *useful* resources that the buyer gets/uses is still $\bar{Q}^B(\alpha_i, \hat{v}_i)$, by considering the following two cases. These are defined depending on whether the allocation partially or fully satisfies buyer $i$'s demand if she submits her true demand $\alpha_i$. In the former case, we prove that her allocation does not change even if a higher demand is submitted given our use of Algorithm 1 in the proof of Proposition 1. In the latter case, Proposition 1's monotonicity property also shows that her new allocation by submitting a higher demand can only be equal to or greater than $\alpha_i$ and the buyer will still only use $\alpha_i$ resources. Thus, her expected utility function satisfies:

$$\bar{U}_c^B(\alpha_i, v_i, \hat{\alpha}_i, \hat{v}_i) - \bar{U}_c^B(\alpha_i, v_i - 1, \hat{\alpha}_i, \hat{v}_i) = \bar{Q}^B(\alpha_i, \hat{v}_i),$$

which is mainly due to the linearity of the utility functions. Combining this equation with (7a), we obtain:

$$\bar{U}_c^B(\alpha_i, v_i, \alpha_i, \hat{v}_i) = \bar{U}_c^B(\alpha_i, \hat{v}_i, \alpha_i, \hat{v}_i) + (v_i - \hat{v}_i)\bar{Q}^B(\alpha_i, \hat{v}_i),$$
$$\bar{U}_c^B(\alpha_i, v_i, \hat{\alpha}_i, \hat{v}_i) = \bar{U}_c^B(\alpha_i, \hat{v}_i, \hat{\alpha}_i, \hat{v}_i) + (v_i - \hat{v}_i)\bar{Q}^B(\alpha_i, \hat{v}_i).$$

Observe that the second terms on the right-hand-side are the same in the above two equations. Thus, $\bar{U}_c^B(\alpha_i, v_i, \alpha_i, \hat{v}_i) \geq \bar{U}_c^B(\alpha_i, v_i, \hat{\alpha}_i, \hat{v}_i)$ is equivalent to $\bar{U}_c^B(\alpha_i, \hat{v}_i, \alpha_i, \hat{v}_i) \geq \bar{U}_c^B(\alpha_i, \hat{v}_i, \hat{\alpha}_i, \hat{v}_i)$.

To show that the buyers have no incentive to over-report their demands, we need to prove $\bar{U}_c^B(\alpha_i, v_i, \alpha_i, v_i) \geq \bar{U}_c^B(\alpha_i, v_i, \hat{\alpha}_i, \hat{v}_i)$ for all $v_i$, $\hat{v}_i$ and $\hat{\alpha}_i > \alpha_i$. Note that the correction payment returned by Algorithm 2 satisfies the incentive compatibility constraints for values that $\bar{U}_c^B(\alpha_i, v_i, \alpha_i, v_i) \geq \bar{U}_c^B(\alpha_i, v_i, \alpha_i, \hat{v}_i)$ when buyers truthfully reveal their demands. Therefore, it is enough for us to prove that $\bar{U}_c^B(\alpha_i, v_i, \alpha_i, \hat{v}_i) \geq \bar{U}_c^B(\alpha_i, v_i, \hat{\alpha}_i, \hat{v}_i)$, which, based on the discussion in the above paragraph, is equivalent to:

$$\bar{U}_c^B(\alpha_i, v_i, \alpha_i, v_i) \geq \bar{U}_c^B(\alpha_i, v_i, \hat{\alpha}_i, v_i), \forall v_i, \hat{\alpha}_i > \alpha_i. \tag{17}$$

First note that we only prove (17) for $\hat{\alpha}_i = \alpha_i + 1$ and the extension to all $\hat{\alpha}_i > \alpha_i$ follows the similar arguments. When $\hat{\alpha}_i = \alpha_i + 1$, buyer $i$ might be assigned with one unit more than what she needs without extra benefit. Thus,

$$\begin{aligned}\bar{U}_c^B(\alpha_i, v_i, \alpha_i + 1, \hat{v}_i) &= \bar{U}_c^B(\alpha_i + 1, v_i, \alpha_i + 1, \hat{v}_i)\\ &\quad -v_i \Pr(X = \alpha_i + 1 | \alpha_i + 1, \hat{v}_i), \forall v_i, \hat{v}_i, \alpha_i,\end{aligned} \tag{18}$$

where $\Pr(X = \alpha_i + 1 | \alpha_i + 1, \hat{v}_i)$ denotes the probability of being assigned with $\alpha_i + 1$ units when a buyer reports $\hat{\alpha}_i = \alpha_i + 1$ and $\hat{v}_i$. Moreover, we note that submitting a higher demand than the true one increase $f_{ij}$ only when the true demand is fully satisfied. Without correction, payment for any extra resource allocation only leads to negative utility, and hence we have:

$$\bar{U}^B(\alpha_i, v_i, \alpha_i + 1, v_i) \leq \bar{U}^B(\alpha_i, v_i, \alpha_i, v_i), \forall v_i, \alpha_i. \tag{19}$$

Next, we start to prove (17) for $\hat{\alpha}_i = \alpha_i + 1$ formally using (18) and (19). Similarly, let $\tau$ denote the minimum value with nonzero correction $\bar{g}(\alpha_i+1, \tau)$. For $v_i < \tau$, we can prove $\bar{U}_c^B(\alpha_i, v_i, \alpha_i+1, v_i) \leq \bar{U}_c^B(\alpha_i, v_i, \alpha_i, v_i)$ as follows:

$$\begin{aligned}\bar{U}_c^B(\alpha_i, v_i, \alpha_i + 1, v_i) &\overset{(g)}{=} \bar{U}^B(\alpha_i, v_i, \alpha_i + 1, v_i)\\ &\overset{(h)}{\leq} \bar{U}^B(\alpha_i, v_i, \alpha_i, v_i) \overset{(i)}{\leq} \bar{U}_c^B(\alpha_i, v_i, \alpha_i, v_i),\end{aligned}$$

where (g) is because $\bar{g}(\alpha_i + 1, v_i) = 0$, (h) is derived by (19) and (i) is due to non-negative correction.

Now, given $\bar{U}_c^B(\alpha_i, v_i, \alpha_i + 1, v_i) \leq \bar{U}_c^B(\alpha_i, v_i, \alpha_i, v_i)$ for $v_i < \tau$, we can further prove $\bar{U}_c^B(\alpha_i, \tau, \alpha_i + 1, \tau) \leq \bar{U}_c^B(\alpha_i, \tau, \alpha_i, \tau)$ as follows:

$$\begin{aligned}&\bar{U}_c^B(\alpha_i, \tau, \alpha_i{+}1, \tau)\\ &\overset{(j)}{=} \bar{U}_c^B(\alpha_i{+}1, \tau, \alpha_i{+}1, \tau) - \tau \Pr(X{=}\alpha_i{+}1 | \alpha_i{+}1, \tau)\\ &\overset{(k)}{=} \bar{U}_c^B(\alpha_i{+}1, \tau, \alpha_i{+}1, \tau{-}1) - \tau \Pr(X{=}\alpha_i{+}1 | \alpha_i{+}1, \tau)\\ &\overset{(l)}{\leq} \bar{U}_c^B(\alpha_i{+}1, \tau, \alpha_i{+}1, \tau{-}1) - \tau \Pr(X{=}\alpha_i{+}1 | \alpha_i{+}1, \tau{-}1)\\ &\overset{(m)}{=} \bar{U}_c^B(\alpha_i, \tau, \alpha_i{+}1, \tau{-}1)\\ &= \bar{U}_c^B(\alpha_i, \tau{-}1, \alpha_i{+}1, \tau{-}1) + \bar{Q}^B(\alpha_i, \tau{-}1)\\ &\leq \bar{U}_c^B(\alpha_i, \tau{-}1, \alpha_i, \tau{-}1) + \bar{Q}^B(\alpha_i, \tau{-}1)\\ &= \bar{U}_c^B(\alpha_i, \tau, \alpha_i, \tau{-}1) \overset{(n)}{\leq} \bar{U}_c^B(\alpha_i, \tau, \alpha_i, \tau),\end{aligned}$$

where (j) and (m) are derive by (18), (k) is derived by (13) and (n) uses the AIC of function $\bar{U}_c^B$ when truthfully reporting demand. (l) is due to the monotonicity property of $\Pr(X = \alpha_i + 1 | \alpha_i + 1, \hat{v}_i)$ in $\hat{v}_i$, which can be derived by the monotonicity property of $\sum_{j \in \mathcal{S}(i)} f_{ij}$ in Proposition 1.

Then, by using mathematical induction, we can finally prove $\bar{U}_c^B(\alpha_i, v_i, \alpha_i, v_i) \geq \bar{U}_c^B(\alpha_i, v_i, \hat{\alpha}_i, v_i)$ for all $v_i$ and $\hat{\alpha}_i = \alpha_i + 1$ as in (17). Therefore, we can conclude that buyers have no incentive to over-report their demands. The proof of the over-reporting case for buyers is completed.

The proof for sellers follows by similar arguments.